%% file: main.tex
\def\year{2021}\relax
\documentclass[letterpaper]{article} 
\usepackage{aaai21}  
\usepackage{times}  
\usepackage{helvet} 
\usepackage{courier}  
\usepackage[hyphens]{url}  
\usepackage{graphicx} 
\usepackage{enumerate}
\usepackage{xspace}
\usepackage{graphicx}  
\usepackage{natbib}
\usepackage{caption}
\usepackage{booktabs}
\usepackage{tikz}
\newcommand*\circled[1]{\tikz[baseline=(char.base)]{
            \node[shape=circle,draw,inner sep=0.8pt] (char) {#1};}}

\usepackage{todonotes}

\title{\modelname: Zero-shot Automatic Multiple-Choice Question Generation\\for Skill Assessments}

\author{
    Eric Li\thanks{Equal contribution.}\textsuperscript{\rm 1}, 
    Jingyi Su\footnotemark[1]\textsuperscript{\rm 1},
    Hao Sheng\textsuperscript{\rm 2}, 
    Lawrence Wai\textsuperscript{\rm 1}\\ 
    {\normalfont 
    \textsuperscript{\rm 1}Chegg, Inc., 
    \textsuperscript{\rm 2}Stanford University}\\
    {\normalfont 
    \{eli, jsu, lwai\}@chegg.com, haosheng@stanford.edu} 
}
 
\begin{document}

\input{macro}

\maketitle

\begin{abstract}
Multiple-choice questions (MCQs) offer the most promising avenue for skill evaluation in the era of virtual education and job recruiting, where traditional performance-based alternatives such as projects and essays have become less viable, and grading resources are constrained. 
The automated generation of MCQs would allow assessment creation at scale. Recent advances in natural language processing have given rise to many complex question generation methods. However, the few methods that produce deployable results in specific domains require a large amount of domain-specific training data that can be very costly to acquire. 
Our work provides an initial foray into MCQ generation under high data-acquisition cost scenarios by strategically emphasizing paraphrasing the question context (compared to the task). In addition to maintaining semantic similarity between the question-answer pairs, our pipeline, which we call \modelname, consists of only pre-trained models and requires no fine-tuning, minimizing data acquisition costs for question generation. \modelname successfully outperforms other pre-trained methods in fluency and semantic similarity. Additionally, with some small changes, our assessment pipeline can be generalized to a broader question and answer space, including short answer or fill in the blank questions.
\end{abstract}

\input{tables/table1}

\section{Introduction}
The outbreak of the COVID-19 pandemic has initiated an acceleration in the shift towards virtual education and job recruiting. There are increased needs for domain-specific skill assessments with feedback and grading that can be provided remotely.
Assessments with MCQs stand out as they can be automatically graded very easily. 
When deployed, they are generally preferred due to fast feedback, flexible testing schedules, and scalable assessment administration~\cite{mccoubrie2004improving}, compared to open-ended assessments such as essays and project-based evaluation. 
Moreover, it is often possible to associate each multiple-choice question with the specific skill and domain it evaluates, and hence detailed feedback can provided.
In Table~\ref{table: mcq}, we show some example questions together with the domains and question types.

Despite its popularity, a high-quality MCQ is costly and time-consuming to generate~\cite{farley1989multiple}. In addition, there are two unique challenges when using MCQs to evaluate a candidate's skills in a given domain, both of which can be mitigated by automated MCQ generation. The first is the possibility that the candidate guesses the correct answer~\cite{kerkman2014challenging}. Statistically, increasing the number of distractor answers will reduce the likelihood of guessing correctly. However, it is costly and extremely difficult to generate appropriate distractor answers~\cite{tarrant2009assessment, haladyna2002review}. As a result, the more feasible approach to alleviate the impact of guessing is to include multiple questions that test the same skill.
The other challenge is how to enable multiple attempts and prevent cheating when the assessment is distributed to a large number of assessment takers~\cite{harmon2010assessment}. This issue can be solved by maintaining a pool of questions which are syntactically and contextually different, but are semantically similar -- i.e., different questions that test the same domain specific knowledge with the same difficulty level. 
Automatic generation of high-quality multiple-choice questions for skill assessments would thereby minimize the impact of guessing and cheating. In addition, it would enable large-scale distribution of assessments and allow each assessment taker multiple attempts at a relatively low cost.

Our work is an initial attempt at high-quality multiple-choice question generation for skill assessments across domains. We introduce \modelname ({\bf A}ssessment {\bf Gen}eration {\bf T}ool with {\bf Zero} fine-tuning), a zero-shot automatic MCQ generation network for skill assessments. With transfer-learning techniques, our method is able to produce high-quality MCQs with no training examples. As a result, this pipeline can be deployed in production for real-time question generation where the output question-answer pairs maintain the same meaning as that of the input question. This feature allows the generated questions to be used for large-scale distribution without concern about normalizing question difficulty. Additionally, our question generation pipeline can be generalized to other question and answer formats, such as fill in the blank questions and true or false questions, via small changes in pipeline hyperparameters, which define input and output format.

In the rest of this article, we first review related works in automated question generation and text-to-text frameworks. We then describe the dataset we used and the architecture of our zero-shot question generation pipeline. We conclude the discussion by evaluating the change of question difficulty, fluency, syntactic difference, and semantic similarity.

\section{Related Work}
\subsubsection{Question Generation and Paraphrasing.}
Recent developments in automated question generation have made many improvements in question complexity and diversity through various system designs~\cite{aqg_review}. 
{\it Syntax-based Methods}~\cite{syntax_qg} use syntactical features of a statement to rephrase it into a question. 
Unfortunately, the outputs generated by these methods are relatively low quality in terms of readability and the variety of questions generated.
Given a context paragraph, {\it Context-based Methods}~\cite{rajpurkar2016squad, du2017learning} are able to generate questions together with the correct answer. The state of the art neural network models have demonstrated their capability in generating reading comprehension questions~\cite{nqg_review, nqg_overview}.
However, such context paragraphs are often not included in the corpus and sometimes are tested as background knowledge themselves. 
{\it Template-based Methods} have performed extremely well on the level of semantic similarity~\cite{domain_qg}, difficulty~\cite{grossman2019mathbot}, and question diversity~\cite{neural_template} for domain-specific tasks. However, current methods rely heavily on a template library, which can be built either manually or automatically through the use of knowledge graphs~\cite{template_gen_knowledgegraph}. 
Though these methods appear to be robust in specific domains (e.g., Biology~\cite{biology_ontology}), they are not broadly applicable in high data-acquisition cost scenarios where knowledge graphs are expensive to compute.

Our work is also closely related to {\it Question Paraphrasing}. The simplest version of question paraphrasing is rule-based word replacement. This can come in the form of named entity recognition and replacement or specific word selection and replacement~\cite{mckeown1980paraphrasing,meteer1988strategies}. More recent state-of-the-art question paraphrasing methods achieve end to end question generation, including the latent bag-of-words~\cite{pg_latentbow}, the reinforcement learning approach~\cite{pg_rl}, and the encoder-decoder approach~\cite{edd-lg}. 
To generate high-quality questions in a specific domain, these methods would require a large number of paired questions for training (e.g., Quora duplicated question pairs~\cite{quora_duplicated_questions}), which can again be costly.

\subsubsection{Text-to-text Neural Networks.}
First adopted by machine translation~\cite{bahdanau2014neural, wu2016google}, a text-to-text (or encoder-decoder) neural net framework has been used in a variety of natural language processing (NLP) tasks such as linguistic acceptability~\cite{warstadt2019neural}, sentence similarity~\cite{mueller2016siamese}, and document summarization~\cite{tan2017abstractive, zhang2019hibert}.  
Recent success in unifying all these tasks with a single text-to-text framework has been ignited by Transformer-like models such as GPT~\cite{radford2018improving}, ELMo~\cite{peters2018deep}, BERT~\cite{bert} and Reformer~\cite{kitaev2020reformer}. 
When pretrained on abundantly-available unlabeled text data with a self-supervised task, such models have led to a new wave of state-of-the-art results in NLP, including question generation~\cite{bert_qg}.
In this paper, Text-To-Text Transfer Transformer (T5)~\cite{T5} is used as our primary backbone model for question generation. We also use the pretrained BERT model, and FitBERT~\cite{havens2019fitbert} to identify and replace the named entities, respectively.

\section{Data}
Our primary dataset of interest is a proprietary dataset created for skill assessment as a part of job recruitment process. Our dataset is composed of less than a thousand multiple-choice questions, which fall into one of eleven domains: Statistics, SQL, Excel, Python, Data Visualization, Machine Learning, Business Analytics, Algorithms, Data Structures and Soft Skills. Each data point incorporates an assessment question, corresponding answer choices, the correct answer, and the associated domain as well as the difficulty and proficiency levels(not used in this paper). To ensure question quality, the data collection and quality testing efforts were performed over the period of a few months. Each question is generated through a human-interactive pipeline involving domain experts, and evaluated by hiring managers for its correctness and appropriateness as job skill assessment. We augment this dataset by manually labeling each question with its corresponding question type (application, concept, or calculation) which we describe further in the question classification subsection of the Pipeline Architecture Section. 20\% of the questions in the dataset are application questions, 5\% are calculation questions and the rest are concept questions.  A sample from our dataset is shown in Table~\ref{table: mcq}. Our goal is to generate questions that are syntactically different but semantically similar to the questions in the dataset to support and supplement skill assessments.

\section{Pipeline Architecture}
The most fundamental requirement for a question generation model for assessment creation is the quality of generated questions. The generated questions should not only be related to the assessment domain, but should also be paired with the correct answer and high-quality distractor answers. As mentioned in previous sections, one method fit all models fail to achieve the high question quality standard in every domain unless a large amount of data is available. To achieve the high-quality required for the assessment question generation task, we first devise a robust question classification method to distinguish different categories of questions, and then develop a separate question generation pipeline. Figure 1 gives a high-level architecture overview of our pipeline. 

\begin{figure}[t]
    \includegraphics[width=\linewidth]{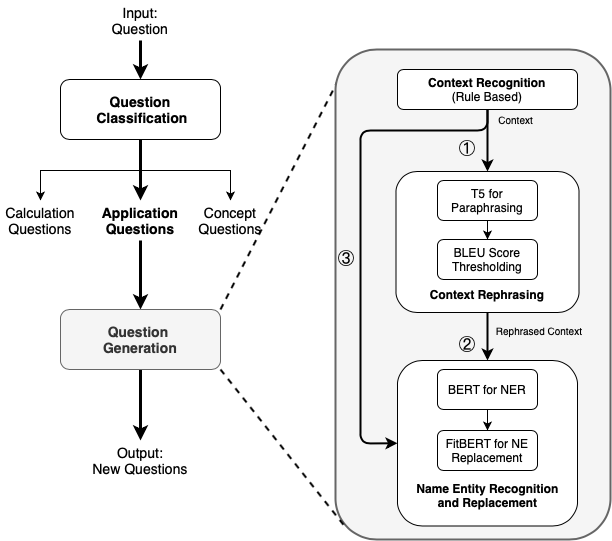}
    \caption{Question generation pipeline with 2 major components: question classification and question generation. The question generation component includes context recognition, context rephrasing and name entity recognition and replacement sub-components to generate new questions. Our default flow for question generation is \circled{1} and \circled{2} if both context rephrasing and name entity recognition can provide good results; otherwise, \circled{1} only if name entity recognition is impossible and \circled{3} only if context rephrasing doesn't meet the BLEU score thresholding requirement. More details are in the Pipeline Architecture Section.}
    \label{fig:full_pipeline}
\end{figure}

\subsection{Question Classification}
Aside from question classification from a learning stand-point~\cite{q_types}, there is no generally accepted question categorization for question generation tasks. This makes it difficult to develop a broad approach to MCQ generation due to the highly variant nature of questions. We describe a categorization system specific to MCQ assessments and the question/answer syntax as follows (illustrated with example questions in Table~\ref{table: mcq}):
\begin{itemize}
    \item \textbf{Application questions}: these questions describe scenarios that require domain specific skills to solve. These questions contain two parts: the first part provides context for the question, the second part is the actual question that needs to be answered.
    \item \textbf{Calculation questions}: these questions ask for numerical answers that are found by applying specific formulas to the values provided in the question. 
    \item \textbf{Concept questions}: these questions are centered on one or multiple domain concepts and terminologies directly. Answers are generally quite short.
\end{itemize}

This categorization stems from each question type's ability to test high-level applied knowledge of the domain in question. Application questions are the best best way to evaluate complex understanding of a domain, concept questions are more recall based, while calculation questions are somewhere in between ~\cite{app_mcq}. This specific categorization is not necessarily applicable in all MCQ generation. That being said, this categorization approach provides a valuable framework for developing subsequent question generation pipelines in order to make each pipeline more generalizable.

Our classification model is a simple logistic regression with a bag of words feature set as well as a few custom features such as number of numeric tokens, length of question, and average length of answer choices.

Multiple-choice questions and their efficacy in learning and skill evaluation have been researched extensively. We observe that questions with a meaningful stem, i.e. the body of text preceding the answer choices, and a complex question component (instead of "which of the following is true") are the most effective at consistently testing high-level skills~\cite{assessment_tips}. This question format matches exactly with our definition of application questions. As a result, to facilitate the generation of assessments for high-level skill evaluation, we choose to focus on application questions over concept or calculation questions. Concept questions lack the necessary depth and breadth of concept coverage for high-level skill evaluation~\cite{good_mcq}. Calculation questions test applied skill well, and are thus valuable. That being said, generation of calculation questions can be rather trivially solved using simple syntactic methods which change the numbers within the question and then apply these changes to the correct answer. As a result, given the value of application questions and their sparsity in our current dataset, we choose to focus on application question generation.

A question generation model is developed for the application question type. The generated questions inherit the original questions' answer options while maintaining semantic similarity and question difficulty.

\subsection{Question Generation}
As shown in Figure~\ref{fig:full_pipeline}, question generation has three major components: context recognition, context rephrasing, and name entity recognition and replacement. Details of each components will be discussed below. New question answer pairs can be generated from context rephrasing alone, named entity recognition and replacement alone, or a combination of both.
Our pipeline will generate new questions with the combined methods by default. In more complex scenarios where high-quality context rephrasing results are not available or there does not exist a replaceable named entity in the original question, this pipeline ensures the generation of new questions with at least one of its components.

\subsubsection{Context Recognition.}
We split each question into two parts, the \textit{context} and the \textit{task}. The context provides background information for the question. The task is the actual question, specified to begin with a WH-phrase (who, what, when, where, why, how).
Suppose we have the following question (question 4 in Table~\ref{table: example}). Note that we omit the answer options for brevity. The first sentence is the context and the second is the task.
\begin{quote}
    Robert and Annie begin arguing during a meeting about how to prepare a presentation. (\textit{context}) What is the first thing they should do to resolve this conflict? (\textit{task})
\end{quote}

In the following steps for question generation, the pipeline focuses on generating the context while keeping the task the same to ensure the answers associated with said question continue to make sense.
\input{tables/table2}
\subsubsection{Context Paraphrasing.}
Question paraphrasing is performed with a pre-trained T5 model fine-tuned for paraphrase generation. Fine-tuned pre-trained paraphrase generation models can achieve good readability and semantic similarity even with very little data. However, the nature of MCQs requires the questions to be precisely matched with the correct answers, and small changes in the question can change the answer completely. Below is an example of the question shown above after being rephrased by fine-tuned T5.
\begin{quote}
    \textit{I know Robert and Annie are arguing during a meeting, what should they do?}
\end{quote}
The generated question does not have the same meaning as the original. Given that we have no precise method to generate correct answers and suitable distractor answers, this generated question cannot be used for assessment purposes.

To combat this, we employ paraphrasing methods only on the context of each question, thereby ensuring that the task in the question stays intact such that the same answers still apply.

Additionally, pre-trained paraphrasing models can generate unrelated paraphrases. For example, considering the original question above, the context rephrasing can generate
\begin{quote}
    \textit{As Bob and Annie argue about preparing a presentation, they start arguing.}
\end{quote}
In order to resolve this, we employ an additional thresholding layer using BLEU-4~\cite{bleu} score to remove vastly different or extremely similar paraphrases.

Following BLEU score thresholding, we perform text based edits to remove certain artifacts such as extra sentences, extra spaces, and misplaced punctuation.

\subsubsection{Named Entity Recognition and Replacement.}
Questions with the same semantic meaning can also be generated by replacing certain types of named entities. We fine-tuned a pre-trained BERT model for named entity recognition (NER) with the CoNLL-2003 dataset~\cite{conll_dataset}. The four entities it recognizes are names, geopolitical entities, locations, and miscellaneous. Miscellaneous entities include a variety of terms ranging from technical terms (such as \textit{Python} and \textit{SQL}) to languages (such as \textit{Greek} and \textit{German}). This variety made it difficult to manually generate possible replacements for entities classified as miscellaneous. To deal with this, we searched for the most likely replacement entities for the named entity. For each entity, in addition to considering the default manually generated replacements suggested, we searched for similar entities using pre-trained word embeddings and add these to the list of potential replacements. 

After developing a list of potential replacements for each replaceable named entity, we apply FitBERT to search for the best replacement. After these replacements are recommended, the model then samples various combinations of these recommended replacements to replace the named-entities in the original question. If named entities that are changed in the context appear in the task, those that appear in the task will be modified as well.

\section{Evaluation}
We compare our pipeline with a pre-trained T5 model adapted for paraphrase question generation. We used two different forms of evaluation: human evaluation and seq2seq machine translation evaluation scores. These are the standard methods by which question generation models are evaluated~\cite{aqg_review} as human evaluation provides the most applicable results while machine translation scores provide the best results for comparison.

The pre-trained T5 model was fine-tuned on the Quora dataset with a beam search layer to find the best paraphrased questions. The specific hyperparameters associated with our pipeline are shown below:
\begin{itemize}
    \item BLEU-4 thresholding for context paraphrasing is  between a minimum of 0.23 and maximum of 0.8.
    \item Pre-trained word embeddings for replaceable part selection are GloVe embeddings~\cite{pennington2014glove} trained on the Wikipedia corpus with 6 billion tokens.
    \item The GloVe embedding search space was constrained to the 5 most similar tokens. We limit by token count as distance in vector space is harder to optimize. 
\end{itemize}

104 questions were randomly selected from our dataset as the original questions for question generation and the generated questions are used for evaluation. These questions span a variety of domains including machine learning, statistics, soft skills, data visualization, SQL, and others.
\input{tables/table3}
Human evaluation was done with domain experts. We evaluated the performance of the generated questions using the following four metrics (results are in Table~\ref{table: human_results}):
\begin{itemize}
    \item \textbf{Question Difficulty:} the difficulty of answering the generated question relative to the original question. Negative values mean that the generated questions is less difficult while positive values mean that the generated question is more difficult.
    \item \textbf{Fluency:} grammatical accuracy of the  question and how fluently the generated question is phrased. Higher values represent better fluency.
    \item \textbf{Syntactic Difference:} the structure and word choice differences between the generated and original question. Higher values mean more different in wording.
    \item \textbf{Semantic Similarity:} whether the generated question retains the same meaning as the original question or not. Higher values mean more similar in meaning.
\end{itemize}
Each of these metrics is rated from 1-5 (where 5 is the best score) with the exception of question difficulty which is rated from -2 to 2 (where 0 is the best score).

The second evaluation method involved comparing our generated data to a dataset generated by subject-matter experts including original and rephrased questions. We compare the rephrased questions from experts to the questions generated by our pipeline by comparing the difference between BLEU~\cite{bleu}, METEOR~\cite{meteor}, ROUGE~\cite{rouge}, and CIDER score~\cite{cider} of the rephrased question and the original. Given that we use human evaluation as a metric for semantic similarity, our goal is not to produce the most syntactically similar paraphrase, but rather produce the most syntactically different paraphrases which also maintain the semantic meaning. As a result, we are looking to minimize machine translation score (instead of maximize). This is in contrast to the traditional usage of machine translation metrics.

The question classifier is evaluated separately using accuracy on an evaluation subset of our data. 

Note, we exclude advanced paraphrase question generation models from our comparison and evaluation analysis as they performed poorly out of the box on our dataset. 

\section{Results and Discussion}
\subsection{Question Classification}
The question classifier performs extremely well on application questions, achieving 97.67\% precision and 91.3\% recall.
Detailed results can be found in Figure~\ref{fig: app_confusion_matrix} in the appendix.
Note that our classifier was optimized for application question classification. As a result, it performs extremely well in this area alongside concept question classification. The precision for calculation questions is quite low, however this may be attributed to a low number of calculation questions in the actual dataset. A robust question classifier is key to the performance of this pipeline.

\subsection{Question Generation}
\input{tables/table4}
\subsubsection{Human Evaluation.}
Table~\ref{table: example} shows the results of human evaluation of questions generated by the full \modelname pipeline, its separate components, and the pre-trained T5 model. Note that not every question included NER replacement; thus the context paraphrasing and NER replacement scores are not the same as the full \modelname scores. It is also important to note that T5 paraphrasing failed to generate a result for one of the questions while our pipeline successfully paraphrased all of the questions.

Overall, our pipeline performed extremely well in terms of fluency and semantic similarity while maintaining question difficulty. They are not as syntactically different as T5 paraphrased questions, however this may be attributed to the propensity for T5 to produce either near exact replicas of original questions or questions that change the meaning of the question.

Table~\ref{table: example} shows some examples of questions generated by \modelname and by the T5 model. The first example shows the poor syntactic difference between the generated question from T5 and the original question (the only difference is that "what is" is transformed to "what's"). In contrast, \modelname produces a much more different result while maintaining question meaning.

The second example indicates how T5 generated a completely different question. Though this would perform well in terms of syntactic difference, the meaning of the question is completely different. On the other hand, \modelname generates a less syntactically different question while maintaining the question meaning, which is the most important consideration.

In addition to comparing our pipeline with the pre-trained T5 model, we also evaluate the performance of individual components of our pipeline to compare how each method performs individually and how they perform when combined. As expected, the combined method provides the greatest syntactic difference. Additionally, it is able to generate the most human-readable questions. However, both of these are at the expense of semantic similarity.

\subsubsection{Translation Score.}
The results of machine translation scores are shown in Table~\ref{table: translation-score}. We compare our pipeline to human and T5 generated questions. Human-generated questions end up being the most different by far. That being said, \modelname significantly outperforms T5 paraphrasing in three out of the four metrics while performing similarly well on BLEU-4. This would indicate that \modelname produces the most syntactically unique generated questions.

\section{Conclusion and Future Work}
We have developed a high-quality multiple-choice question generation pipeline that requires no additional training data for assessment creation. \modelname can generate a large number of new questions given a single input question while maintaining semantic similarity for assessments. This pipeline is applicable to question types besides MCQ as the only required inputs are a question and its answer(s), and subsequently addresses generating a large number of precise MCQ assessments.

In the future, we hope to investigate robust automated methods for concept question generation via automated knowledge graph creation and template generation using a pre-existing assessment.


\bibliography{reference}

\newpage
\input{appendix}
\end{document}

%% file: macro.tex
\newcommand{\modelname}{AGenT Zero\xspace}

%% file: tables/table1.tex
\begin{table*}[t]
\begin{center}
\begin{tabular}{ p{0.8cm}|p{5.5cm}|p{5.5cm}|p{1.5cm}|p{1.5cm} }
 \toprule
 \textbf{Index} & \textbf{Question} & \textbf{Answer Choices} & \textbf{Domain} & \textbf{Question Type}\\
 \midrule \vspace{.5em}1& \vspace{.5em}
 You are building a SVM.  The dataset has N samples  The number of features is D.  N $<$ D.  What kernel function should you start with? &
 \begin{enumerate}[(a)]
     \item RBF kernel
     \item Polynomial kernel
     \item Quadratic kernel
     \textbf{\item Linear kernel}
 \end{enumerate} & \vspace{.5em} Machine \quad Learning & \vspace{.5em} Application\\
 \midrule \vspace{.5em}2&
 \vspace{.5em} What SQL command should be used for performance to delete all records from a table without removing the table object? &
 \begin{enumerate}[(a)]
     \item DELETE
     \item UPDATE
     \textbf{\item TRUNCATE}
     \item DROP
 \end{enumerate}
 & \vspace{.5em} SQL & \vspace{.5em} Concept\\
 \midrule \vspace{.5em}3&\vspace{.5em} Let's say that you sell a washing machine for \$500, which cost you \$300 to make.  What is your gross margin? &
 \begin{enumerate}[(a)]
    \item \$200
    \textbf{\item 40\%}
    \item 167\%
    \item 67\%
 \end{enumerate}&\vspace{.5em} Business Analytics & \vspace{.5em} Calculation\\
 \midrule \vspace{.5em}4&\vspace{.5em}
 Robert and Annie begin arguing during a meeting about how to prepare a presentation. What is the first thing they should do to resolve this conflict? &
 \begin{enumerate}[(a)]
    \item Look for common ground
    \item Ask another meeting attendee to provide a neutral opinion
    \textbf{\item Listen carefully to make sure they understand the problem}
    \item Temporarily change the subject to something less argumentative
 \end{enumerate}&\vspace{.5em} Soft Skills & \vspace{.5em} Application\\
 \bottomrule

\end{tabular}
\end{center}
\caption{Examples of multiple choices questions, their associated skills and their question types. These examples are from our data set used for pipeline evaluation. The {\bf correct answers} are marked in bold. }
\label{table: mcq}

\end{table*}

%% file: tables/table2.tex
\begin{table*}[!htbp]
\centering
\begin{tabular}{l | c c c c}
\toprule
\textbf{Model} & \shortstack{Change in\\Difficulty} & Fluency & \shortstack{Syntactic\\Difference} & \shortstack{Semantic\\Similarity}\\
\midrule
Pre-trained T5 &-0.11 &4.10&\textbf{2.72}& 3.51\\
\textbf{\modelname (full pipeline)}& \textbf{0.01} & \textbf{4.34} & 2.28 & \textbf{4.33}\\
\midrule
\modelname (context paraphrasing only) & -0.02 & 4.32& 2.27&4.17\\
\modelname (NER replacement only) &\textbf{0.00}& 4.32&2.24&\textbf{4.57}\\
\modelname (context paraphrasing and NER replacement) &0.05&\textbf{4.37}&\textbf{2.34}&4.08\\
\bottomrule
\end{tabular}
\caption{Human evaluation results of our pipeline, components of our pipeline, with comparison of the pre-trained T5 model for paraphrased question generation. Change in difficulty is scored from -2 to 2 where 2 represents an increase in difficulty while -2 is a decrease in difficulty, thus making 0 the best score. The other values are evaluated from 1 to 5 where the higher values represent better scores.}
\label{table: human_results}
\end{table*}

%% file: tables/table3.tex
\begin{table*}[!htbp]
\centering
\begin{tabular}{p{1cm}|p{5cm}|p{5cm}|p{5cm}}
 \toprule
 Index & Original Question & Question Generated by T5 & Question Generated by AG-Net \\
 \midrule
 1&Given a method declared as public static void doStuff(char ch) inside a class called MyChar, what is the correct way to call the method from a different class?&Given a method declared as public static void doStuff(char ch) inside a class called MyChar, what's the correct way to call the method from a different class?&If a method is declared as public static void doStuff(char ch) inside a class called VarName, what is the correct way to call the method from a different class?\\
 \midrule
 2&The lifetime of a certain brand of lightbulbs is normally distributed with a mean of 1,500 hours and a standard deviation of 50 hours. What is the probability that a randomly chosen lightbulb will last at most 1,560 hours? & Is it probable that the light bulb will last more than 1.5 hrs? & The life of a given brand of light bulbs is spread in a normal distribution where the mean is 1,500 hours and the standard deviation is 50 hours. What is the probability that a randomly chosen light bulb will last at most 1,560 hours?\\
 \bottomrule
\end{tabular}

\caption{Examples of questions (omitted answer options) generated by AG-Net and by the pre-trained T5 model. Note that we omit answer choices as the focus is on comparing the quality and meaning of generated questions.}
\label{table: example}
\end{table*}

%% file: tables/table4.tex
\begin{table*}[!htbp]
\centering
\begin{tabular}{ p{3cm} | c c c c }
 \toprule
 \textbf{Model} &BLEU-4&METEOR &ROUGE-L&CIDEr \\
 \midrule
 \modelname (ours) &56.93&\textbf{37.93}&\textbf{55.68}&\textbf{3.88}\\
 Pre-trained T5 &\textbf{56.08}&41.37&69.86&4.53\\
 Human generated & 6.81&11.35&23.06&0.21\\
 \bottomrule
\end{tabular}
\caption{Machine translation scores on paraphrased dataset with human translations shown as a score minimum. Note we are attempting to minimize machine translation scores.}
\label{table: translation-score}
\end{table*}

%% file: appendix.tex
\appendix
\section*{Appendix}
\input{figures/app_figure1}

%% file: figures/app_figure1.tex
\begin{figure}[ht!]
    \centering
    \includegraphics[scale=.5]{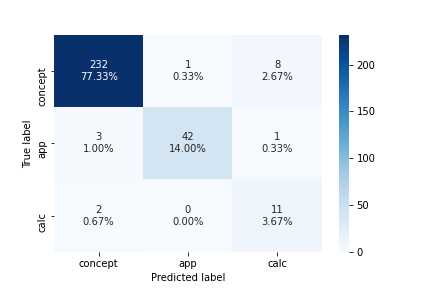}
    \caption{Question classifier results visualized as a multi-class confusion matrix heatmap.}
    \label{fig: app_confusion_matrix}
\end{figure}